\documentclass{jpsj2}

\title{%
 Quantum tunneling in $^{277}$112 and its $\alpha$-decay chain
}

\author{%
C. Samanta$^{1,2}$\thanks{E-mail address: chhanda.samanta@saha.ac.in; csamanta@vcu.edu},
D.N. Basu$^3 $\thanks{E-mail address: dnb@veccal.ernet.in}
and P. Roy Chowdhury$^1$\thanks {E-mail address: partha.roychowdhury@saha.ac.in}
}

\inst{%
$^1$ Saha Institute of Nuclear Physics, 1/AF Bidhan Nagar, Kolkata 700 064, India\\
$^2$ Physics Department, Virginia Commonwealth University, Richmond, VA 23284-2000, U.S.A.\\
$^3$ Variable  Energy  Cyclotron  Centres, 1/AF Bidhan Nagar, Kolkata 700 064, India 
}

\recdate{\today}

\abst{%
The $\alpha$-decay half lives of nuclei in the decay from element $^{277}112$ are calculated in a WKB framework using DDM3Y interaction and experimental Q-values. Theoretical estimation of half lives in the same quantum tunneling model, using Q-values from the mass formula of Muntian-Hofmann-Patyk-Sobiczewski, are also presented. Calculated results furnish corroborating evidence for the experimental findings at RIKEN and GSI. Certain discrepancies indicate necessity of a better mass formula. Further experimental data with higher statistics would also be useful.}

\kword{%
SHE; Alpha decay; Half life; Fusion reaction; Mass formula; DDM3Y.
}

\begin{document}
\maketitle

\section{Introduction}

Advent of radioactive ion beam facility in Japan has added a new dimension to the research on superheavy elements (SHE). This facility has not only led to interesting new discoveries~\cite{mo104,mo1041,mo204,mo107}, but also provided an alternative set up for reconfirmation of the existing ones and can explore theoretical predictions~\cite{ta98,ta00}. Earlier, signatures of several superheavy nuclei from Z~=~112~-~117 and 118 were observed at GSI, Germany and JINR-FLNR, Russia~\cite{ho96,ho00,ho02,oga99,oga104,oga204,oga05,oga06}. However, small production cross sections and the associated difficulties in measurements sometimes led to ambiguities and controversies~\cite{ho02}. Therefore, while rigorous searches for new superheavy elements are on, reconfirmation of the existing ones on a different experimental environment is equally important for the official recognition~\cite{bar92,kar01} as well as progress in this field. 

Recently, existence of the superheavy element $^{277}112$, discovered earlier at GSI, Germany~\cite{ho96,ho00,ho02}, has been reconfirmed in an experiment at RIKEN~\cite{mo207}. The observed decay chains from the GSI and RIKEN show slight variation in decay energies as well as decay times. Nevertheless, all the four $\alpha$-decay chains observed by GSI and RIKEN are basically similar, except the Chain 3 of GSI which extends up to $^{257}$No. The $\alpha_5$ and $\alpha_6$ (Chain 3) were observed by the GSI group only. In Ref.~\cite{mo207} average $\alpha$-decay half life values for $^{277}112$, $^{273}110$, $^{269}108$ and $^{265}106$ were presented, while average half life for the $^{261}104$  was computed using both the $\alpha$-decay and spontaneous fission half lives.

In this work, for the $\alpha$-decay chains from $^{277}112$, a comparison between the experimental data of GSI~\cite{ho96,ho00,ho02} and RIKEN~\cite{mo207} and theoretical predictions of $\alpha$-decay half lives with experimental ($Q^{ex}$) and theoretical  ($Q^M$) Q-values are presented. The aim of this work is to check the possible origin of the above mentioned variation in decay energies and decay times in a formalism based on quantum tunneling model along with microscopic potentials which has a firm theoretical footing compared to semi-empirical approaches\cite{po06}. Also, simultaneous comparison between experimental and theoretical Q-values and half lives would provide a clear demarcation of the extent of validity of the theoretical Q-values.

\section{Microscopic potentials and quantum tunneling}

Theoretical half lives of $\alpha$-decay sequence from $^{277}112$ have been computed for the first time in a WKB framework with DDM3Y interaction using Q-values ($Q^M$) from the mass formula of Muntian-Hofmann-Patyk-Sobiczewski~\cite{mu01,mu103,mu203} and compared with the experimental data of GSI and RIKEN. 

\subsection{Effective interaction and double folded potential}

The theoretical $\alpha$-decay half lives have been calculated using the experimental $Q$-values ($Q^{ex}$) extracted from the measured decay  energies~\cite{mo207}. A comparison with the experimental half lives is presented as well. Half lives of parent nuclei $^AZ$, with the charge number Z~=~112 decaying via $\alpha$ emissions are calculated in the WKB barrier penetration framework using microscopic potentials for the $\alpha$ - nucleus interaction~\cite{ba03}. The nuclear potentials have been obtained microscopically by double folding the $\alpha$ and daughter nuclei density distributions with the density dependent M3Y (DDM3Y) effective nucleon-nucleon interaction. The double folding potential, thus obtained, has been utilised for calculating the barrier penetration probability in a quantum tunneling model.

\subsection{Barrier penetrability in WKB framework}
 
The half life of a parent nucleus decaying via $\alpha$ emission is calculated using the WKB barrier penetration probability~\cite{ke35}. The barrier pentrability with DDM3Y interaction is used to provide estimates of  $\alpha$-decay half lives for $Z~= 102 - 112$  $\alpha$-emitters.  Earlier it was shown~\cite{prc06,prc07} that this procedure of obtaining nuclear interaction energy for the $\alpha$ - nucleus interaction is more fundamental in nature and the half lives calculated in this framework is more reliable than by other methods. It was also shown~\cite{cs07} that the theoretical $Q$ values, called $Q^M$, extracted from the  mass formula of Muntian et al. ~\cite{mu01,mu103,mu203} can reasonably reproduce the experimental data on several SHE. 

The barrier penetrability $P$ in the improved WKB~\cite{ke35} framework for any continuous (rounded) potential barrier  is given by,

\begin{equation}
 P = 1/ [1 + \exp(K)]
\label{seqn1}
\end{equation}
\noindent
where $K$ is the action integral~\cite{cs07} and the decay constant $\lambda=\nu P$ where $\nu$ is calculated from $E_v=\frac{1}{2}h\nu$, the zero point vibration energy. The zero point vibration energies used in the present calculations are $E_v$ = 0.1045$Q$ for even-even, 0.0962$Q$ for odd Z-even N, 0.0907$Q$ for even Z-odd N, 0.0767$Q$ for odd-odd parent nuclei and are the same as that described in ref. \cite{Po86} immediately after eqn.(4) which were obtained from a fit to a selected set of experimental data on $\alpha$ emitters and includes the shell and the pairing effects. The half life can thus be obtained from $T_{1/2}=\ln2/\lambda$. It is worthwhile to mention here that some disagreement with the results of the calculations of Ref.~\cite{ma06} with the experimental results of Ref.~\cite{ho00} may be attributed to the use of zero point vibration energy calculated differently and the fitting of the microscopic folded potentials to the Saxon-Woods form whereas use of experimental $Q$ values instead of theoretical ones would have improved results for Z=102, 104, 106 while the rest three for Z=108, 110, 112 would have further deteriorated.

\subsection{Measured $\alpha$-particle kinetic energies and decay $Q$-values}

      The $\alpha$-decay $Q^{ex}$ values for the favored decays have been calculated from the measured $\alpha$- particle kinetic energies $E$ using standard recoil correction and the electron shielding correction in a systematic manner as suggested by Perlman and Rasmussen\cite{pe57}. The decay $Q^{ex}$ value and the  measured $\alpha$-particle kinetic energy $E$ are related by the following expression: 

\begin{equation}
 Q^{ex} = (\frac{A_p}{A_p-4})E + (65.3 Z_p^{7/5} - 80.0 Z_p^{2/5}) \times 10^{-6} ~\rm MeV
\label{seqn2}
\end{equation}
\noindent
where the first term in the right hand side is the standard recoil correction and the second term is an electron shielding correction. $Z_p$ and $A_p$ are the atomic and mass numbers of the parent nucleus. From the measured decay times (T) we computed the half lives $T^{exp}_{1/2} = 0.693 \times T$. The theoretical $Q$-values are calculated using the mass formula using the following expression:  

\begin{equation}
 Q^{M} = M_{parent} -(M_\alpha+M_{daughter}) MeV
\label{seqn3}
\end{equation}
\noindent
where $M_{parent}$, $M_\alpha$ and  $M_{daughter}$, the masses of parent, alpha and daughter nuclei in MeV, respectively, are calculated from Muntian et al.~\cite{mu01,mu103,mu203}. Certainly these theoretical $Q^M$ values are for ground state to ground state transitions. The experimental $Q^{ex}$ values are not necessarily always the same for they may be for transitions from (i) ground state to ground state, (ii) ground state to excited state, (iii) excited state to ground state or (iv) excited state to excited state. Moreover, errors in theoretical $Q^M$ values are also not provided in ref.\cite{mu01,mu103,mu203}. However, the theoretical half life calculations with experimental $Q^{ex}$ values are correct since the theory as such also takes care of decays other than ground state to ground state provided corresponding $Q^{ex}$ values and spin-parities of parent and daughter nuclei are known. 

\subsection{Spin-parity conservation and the centrifugal barrier}

      The spin-parity conservation condition in a decay process is fulfilled if and only if 

\begin{equation}
 {\bf J} =  {\bf J}_1 + {\bf J}_2 + {\bf l},~~~~~~~\pi =  \pi_1.\pi_2.(-1)^l,  
\label{seqn4}
\end{equation}
\noindent
where ${\bf J}$, ${\bf J}_1$ and ${\bf J}_2$ are the spins of the parent, daughter and emitted nuclei respectively, $\pi$,  $\pi_1$ and $\pi_2$ are the parities of the parent, daughter and emitted nuclei respectively, and {\bf l} is the orbital angular momentum carried away in the process. This conservation law, thus, forces a minimum angular momentum to be carried away in the decay process. Consequently, contribution of the angular momentum gives rise to a centrifugal barrier

\begin{equation}
 V_l = \hbar^2 l(l+1) / (2\mu R^2)
\label{seqn5}
\end{equation}
\noindent
where $\mu$ is the reduced mass of the daughter and emitted nuclei system and $R$ is the distance between them. Although spin of $\alpha$ nucleus is zero and parity is even, yet non-zero spins of the parent and daughter nuclei can force a minimum of 5$\hbar$ of angular momentum to be carried away for parent nuclei around atomic number $Z_p$ = 90 \cite{ba03} and for the present case of nuclei with atomic number  $Z_p$ = 112, in some cases it may be even higher. The uncertainties of the angular momentum transfers arise due to non-availability of the spin-parities of the parent and/or daughter nuclei of the SHE.  
 
\section {Calculations and Results} 

      The basic problem of the present theoretical study is that there is no guarantee that the experimentally observed $\alpha$ decay chains proceed from the ground state of the parent nucleus to that of the daughter nucleus. This is the fundamental difficulty of the decay of the odd mass nuclei, especially for the decay of the SHE, where there are only few data available. It is important to mention here that the theoretical estimates of the $Q$-values ($Q^{M}$) extracted from the mass formula of Muntian et al.~\cite{mu01,mu203} and the half lives [T$^{Q^M}_{1/2}$] have been done presuming ground state to ground state decays. However, the theoretical half life [T$^{Q^{ex}}_{1/2}$] calculations with experimental $Q$-values ($Q^{ex}$) take care of this problem since the theory as such also takes care of the decays other than the ground state to ground state provided the corresponding experimental $Q^{ex}$ values are known.  

      Table-I shows that the calculated values are in reasonably good agreement with most of the experimental data. The calculation of alpha decay half lives is extremely sensitive to the choice of $Q$-values of the reaction.
Theoretical decay $Q$-values ($Q^{M}$) extracted from the mass formula of Muntian et al.~\cite{mu01,mu203} are also presented along with the half lives [T$^{Q^M}_{1/2}$] computed in the same framework.  Spontaneous fission (SF) terminates Chain 1 and Chain 2 of the RIKEN data and Chain 4 of the GSI data. The $\alpha_5$ and $\alpha_6$ decay chains, observed by GSI (Chain 3), were not observed by the RIKEN group. In Ref. ~\cite{mo207} the half life value for the $\alpha_5$ decay was deduced by taking average of both spontaneous fission (SF) half lives of RIKEN data and half life values of SF and $\alpha$-decay of GSI data. Thus the assigned half life~\cite{mo207} of $^{261}Rf$ given in the Table-II should not be compared with our calculated $\alpha$-decay half life. 

\begin{table}[t]
\caption{A Comparison between experimental and calculated $\alpha$-decay half lives of nuclei is listed.  The theoretical $Q^{M}$ values  are deduced  from the mass formula of I. Muntian, S. Hofmann,  Z. Patyk and A. Sobiczeski~\cite{mu01,mu103,mu203} using eqn.3. The E and T values are taken from Morita et al.~\cite{mo207} while those for SF and escape events are omitted here. The $Q^{ex}$ values are derived from E using eqn.2 and $T^{exp}_{1/2} = 0.693 \times T$. Theoretical half lives T$^{Q^{ex}}_{1/2}$ and T$^{Q^M}_{1/2}$ are calculated with $Q^{ex}$ and $Q^{M}$ values, respectively, using DDM3Y interaction in a WKB framework. These half lives are calculated for zero angular momentum transfers. Mean half life was assigned by Morita et al.~\cite{mo207} from the experimental half lives.}
\label{t1}
\begin{center}
\begin{tabular}{lllllll} \hline
\hline
  & & RIKEN & & GSI &  &Mean\\ \hline
  & & Chain 1 & Chain 2& Chain 3 & Chain 4 & Half life\\ 
 \hline
&&&&\\
$\alpha_1$&E (MeV)&$11.09(7)$&$11.32(4)$&$11.45(2)$&$11.17(2)$& $^{277}112$ \\ 
                &$Q^{ex}$ (MeV)&$11.300(71)$&$11.534(40)$&$11.666(20)$&$11.381(21)$& \\ 
                &$Q^{M}$ (MeV)&$12.11$& & & & \\ 
                &T (ms)&1.10&1.22&0.28&1.41&\\
                &T$^{exp}_{1/2}$ (ms)&0.76&0.85&0.19&0.98&$0.69^{+0.69}_{-0.23}$  \\
                &T$^{Q^{ex}}_{1/2}$ (ms)&$0.43^{+0.19}_{-0.13}$&$0.13^{+0.02}_{-0.03}$  
                &$0.06^{+0.01}_{-0.01}$ &$0.28^{+0.03}_{-0.03}$ &\\ 
                &T$^{Q^M}_{1/2}$ (ms)&&&&&0.01\\ \hline
&&&&\\
$\alpha_2$&E (MeV)&$11.14(4)$&$11.15(7)$&$11.08(2)$&$11.20(2)$&  $^{273}Ds$ \\ 
                &$Q^{ex}$ (MeV)&$11.352(41)$&$11.362(71)$&$11.291(21)$&$11.413(20)$&  \\ 
                &$Q^{M}$ (MeV)&$11.1$& & & & \\ 
                &T (ms)&0.520&0.0399&0.11&0.31& \\
                &T$^{exp}_{1/2}$ (ms)&0.360&0.028&0.076&0.215&$0.17^{+0.17}_{-0.06}$ \\
                &T$^{Q^{ex}}_{1/2}$ (ms)&$0.082^{+0.018}_{-0.016}$&$0.077^{+0.034}_{-0.023}$    
                &$0.116^{+0.008}_{-0.016}$&$0.060^{+0.006}_{-0.006}$ & \\ 
                &T$^{Q^M}_{1/2}$ (ms)&&&&&0.30\\ \hline
&&&&\\
$\alpha_3$&E (MeV)&$9.17(4)$&$9.25(7)$&$9.23(2)$&$9.18(2)$&  $^{269}Hs$ \\ 
                &$Q^{ex}$ (MeV)&$9.354(41)$&$9.435(71)$&$9.415(20)$&$9.364(20)$& \\ 
                &$Q^{M}$ (MeV)&$9.14$& & & & \\ 
                &T (s)&14.2&0.270&19.7&22.0& \\
                &T$^{exp}_{1/2}$ (s)&9.84&0.19&13.65&$^{a)}$$14.^{+26}_{-6}$&$9.7^{+9.7}_{-3.2}$ \\
                &T$^{Q^{ex}}_{1/2}$ (s)&$2.44^{+0.79}_{-0.57}$&$1.42^{+0.87}_{-0.53}$    
                &$1.62^{+0.25}_{-0.20}$&$2.29^{+0.30}_{-0.33}$ & \\ 
                &T$^{Q^M}_{1/2}$ (s)&&&&&10.7\\ \hline
\hline
\end{tabular}
\end{center}
\medskip
\end{table}

\begin{table}[t]
\caption{Same as Table-I.}
\label{t2}
\begin{center}
\begin{tabular}{lllllll} \hline
\hline
  & & RIKEN  && GSI &  & Mean   \\ \hline
  & & Chain 1 & Chain 2& Chain 3 & Chain 4 & Half life\\ 
\hline
&&&& \\
$\alpha_4$&E (MeV)&$8.71(4)$&$8.70(4)$& $Escape$    & $Escape$     &  $^{265}Sg$ \\ 
                &$Q^{ex}$ (MeV)&$8.888(40)$&$8.878(40)$ &     & &  \\ 
                &$Q^{M}$ (MeV)&$8.63$& & & & \\ 
                &T (s)&23.0&79.9&7.4&18.8&\\
                &T$^{exp}_{1/2}$ (s)&15.9&55.4&5.13&$^{a)}$$9^{+17}_{-4}$&$22^{+22}_{-8}$  \\
                &T$^{Q^{ex}}_{1/2}$ (s)&$12.59^{+4.32}_{-3.52}$&$13.54^{+4.67}_{-3.78}$    
                & & & \\ 
                &T$^{Q^M}_{1/2}$ (s)&&&&&83.8\\ \hline
&&&&\\
$\alpha_5$&E (MeV)& SF   & SF   &$8.52(2)$&SF    &$^{261}Rf$ \\ 
                &$Q^{ex}$ (MeV)&  -   &   &$8.696(21)$&    &  \\ 
                &$Q^{M}$ (MeV)&$8.53$& & & & \\ 
                &T (s)& -   &    &4.7&    & \\
                &T$^{exp}_{1/2}$ (s)& -   &    &$^{a)}$4.2$^{+3.4}_{-1.3}$&    &$^{b)}$$5.3^{+5.3}_{-1.8}$ \\
                &T$^{Q^{ex}}_{1/2}$ (s)&  -  &    
                & $9.34^{+1.50}_{-1.29}$&   & \\ 
                &T$^{Q^M}_{1/2}$ (s)&& 
                & &&31.7\\ \hline
&&&&\\
$\alpha_6$&E (MeV)&  --  &  --  &$8.34(2)$& -- &$^{c)}$$^{257}No$\\ 
                &$Q^{ex}$ (MeV)&   -  &   &$8.514(21)$&    &  \\ 
                &$Q^{M}$ (MeV)&$8.19$& & & & \\ 
                &T (s)& -   &    &15.0&    &    \\
                &T$^{exp}_{1/2}$ (s)&  -  &    &10.4&    & \\
                &T$^{Q^{ex}}_{1/2}$ (s)& -   &    
                & $6.41^{+1.08}_{-0.89}$&   & \\ 
                &T$^{Q^M}_{1/2}$ (s)&& 
                & &&78.0\\ \hline
\hline
\end{tabular}
\end{center}
\medskip
$a)$ From Ref.~\cite{ho02}\\
$b)$ This value was deduced in Ref.~\cite{mo207} taking average of both SF and $\alpha$-decay half lives.\\ 
$c)$$~^{257}No$ has an electron capture (EC) branch as predicted in Ref.~\cite{ho02}
\end{table}
      
For $\alpha_1$ the half lives calculated with $Q^{ex}$ are in reasonable agreement with the experimental half lives of chains 1, 2, 3 and 4, but $Q^M$ under predicts the $T_{1/2}$ values and over predicts the Q values. It is interesting to note that the half lives calculated with the $Q^M$ values are in excellent agreement with the $\alpha_2$ and $\alpha_3$ of Chain 1 of RIKEN data and Chain 4 of GSI data although the $Q^M$ and $Q^{ex}$ values are slightly different ($\approx 0.2$ MeV). But, for the same $\alpha_2$ and $\alpha_3$,   $T_{1/2}$  values with $Q^{ex}$ underestimates Chain 1 by a factor of $\sim$4. For  $\alpha_4$,  $T_{1/2}$  values with $Q^{ex}$ agree  well with the experimental values for chain 1 and within a factor of $\sim$4  for chain 2 while, $T_{1/2}$ values obtained using $Q^M$  is close to only Chain 2 of RIKEN data. The $Q^M$ values over predict half life experimental data of $\alpha_5$ and $\alpha_6$ (Chain 3) of the GSI group, although $Q^{ex}$ and $Q^M$ values differ only by $\approx$ 0.2 MeV and $\approx$ 0.3 MeV respectively. 

Altogether it appears that both the decay energies and decay times need to be measured with higher statistics. It may be noted that the mass formula used here was specifically developed for heavy and superheavy nuclei. The mass formula itself may need some improvement too. But, unless more experimental data with higher statistics for these nuclei are available, one can not totally rule out the $Q$-value and half life predictions obtained with this mass formula. 

\section{Summary and Conclusions}

      In summary, the new experimental data of RIKEN have provided important confirmation of the element $^{277}112$. We have calculated the $\alpha$-decay half lives of $\alpha$ chains from $^{277} 112$  in the WKB framework with DDM3Y interaction which is known to provide good estimates of the experimental data when experimental $Q$-values are used~\cite{prc06,prc07,cs07}. The calculated half lives are extremely sensitive to the $Q$ values, and a small change affects the results significantly. The half lives calculated with experimental $Q$ values are in reasonable agreement with the experimental data of both GSI~\cite{ho96,ho00,ho02} and RIKEN~\cite{mo207}. 

      Theoretical $Q$-values ($Q^M$) are obtained from the mass formula of I. Muntian, S. Hofmann, Z. Patyk and A. Sobiczewski~\cite{mu01,mu103,mu203}. While the agreement with the $\alpha_2$ and $\alpha_3$ chain half lives agree extremely well with the experimental half life data, the same for $\alpha_1$ under predicts (Table-I). In this work, the half lives listed in Table-I are calculated assuming that the orbital angular momentum $l$ between residual daughter nucleus and the $\alpha$ particle is zero. The under prediction is a possible indication of non-zero angular momentum ($l$) transfer which causes higher potential barrier leading to higher alpha-decay half life. In fact, for $\alpha_1$, higher values of $l$-transfer give a better agreement ($T_{1/2}^{Q^M}$= 0.01~ms for l=0,~0.34~ms for l=6 and ~1.23~ms for l=7 ) with the experimental mean half life value ($T_{1/2}^{exp}=0.69^{+0.69}_{-0.23}$) of $^{277}112$. On the other hand, for the $\alpha_4$, $\alpha_5$ and $\alpha_6$, theoretical alpha-decay half lives over predict the experimental ones. Such over prediction can not be accounted by non-zero $l$-transfer. 

In the context of alpha-decay half lives, excellent agreement between the experimental T$^{exp}_{1/2}$=15.9 s and calculated T$^{Q^{ex}}_{1/2}$=$12.59^{+4.32}_{-3.52}$ s obtained using experimental $Q^{ex}$=8.888(40) MeV of $\alpha_4$ (chain 1) delineates applicability of the WKB formalism with DDM3Y interaction. At the same time, the large value of T$^{Q^M}_{1/2}$=83.8 s for $Q^{M}$=8.63 MeV shows that a small variation of Q-value ($\approx$ 0.2 MeV) can cause a large difference in the calculated alpha-decay half lives. Therefore, further accuracy of the mass formula is needed.

It is also noted that both the calculated  $Q^M$-values and half live  (T$^{Q^M}_{1/2}$) values do not always agree with the experimental $Q^{ex}$-values and half live (T$^{exp}_{1/2}$) values simultaneously. While the mass formula itself may need some improvement, to resolve these discrepancies with the theoretical predictions further experimental data with better statistics are desirable.

\end{document}